\documentclass[aps,pra,preprint,groupedaddress,showpacs]{revtex4}

\usepackage{graphicx}% Include figure files
\usepackage{bm}% bold math
\usepackage{amsmath,amssymb}
\usepackage{txfonts}

\begin{document}

%\preprint{}

\title{R-matrix calculation of integral and differential cross sections 
for low-energy electron impact excitations of N$_2$ molecule}


%\author{Motomichi Tashiro and Keiji Morokuma}
\author{Motomichi Tashiro}
\thanks{Present address: Fukui Institute for Fundamental Chemistry,
Kyoto University, Takano-Nishi-Hiraki-cho 34-4, Kyoto 606-8103, JAPAN.}
\email[E-mail:]{tashiro@fukui.kyoto-u.ac.jp}

\author{Keiji Morokuma}
%\thanks{}
\affiliation{Department of Chemistry, Emory University, 1515 Dickey Drive, Atlanta, Georgia 30322, USA.}


\date{\today}

\begin{abstract}
% put some surmarizing sentence.
%
Low-energy electron impact excitations of N$_2$ molecules are 
studied using the fixed-bond R-matrix method based on 
state-averaged complete active space SCF orbitals. 
Thirteen target electronic states of N$_2$ are included 
in the model within a valence configuration interaction 
representations of the target states. 
Integrated as well as differential cross sections of the 
$A^{3} \Sigma_{u}^{+}$,  $B^{3} \Pi_{g}$, $W^{3} \Delta_{u}$, 
${B'}^{3} \Sigma_{u}^{-}$, ${a'}^{1} \Sigma_{u}^{-}$, 
$a^{1} \Pi_{g}$, $w^{1} \Delta_{u}$ 
and $C^{3} \Pi_{u}$ states are calculated and compared with the 
previous experimental measurements. 
These excitations, especially of the higher four states, 
have not been studied enough theoretically 
in the previous literature. 
In general, good agreements are observed both in the 
integrated and differential cross sections. 
However, some discrepancies are seen in the integrated 
cross sections of the $A^{3} \Sigma_{u}^{+}$ 
and $C^{3} \Pi_{u}$ states, especially  
around a peak structure. 
\end{abstract}

%\keywords{}

\maketitle

%\section{}
%\subsection{}
%\subsubsection{}

\section{Introduction}
%motivation/importance
%check details of how e-N2 excitation data is used. 
Electron impact excitation of nitrogen molecules plays an 
important role in atmospheric emission of planets and satellites 
such as the Earth, Titan and Triton. 
For example, excitation of the ${a}^{1} \Pi_g$ state 
and subsequent transitions to the ground ${X}^{1} \Sigma_g^+$ 
state are responsible for the far ultraviolet emissions of the 
Lyman-Birge-Hopfield system which is prominent in 
the airglow of the Earth's atmosphere\cite{SpaceScienceRev.58.1}.  
%
% a and C states excitation
%
%Nitrogen molecules are major constituent of 
%the atomosphere of Earth, Titan and Triton. 
% a bit more specic..
Recently, Khakoo et al.\cite{2005PhRvA..71f2703K} measured 
differential cross sections (DCSs) 
of electron impact excitation of N$_2$ molecule from the ground 
${X}^1 \Sigma^{+}_{g}$ state to the 8 lowest excited electronic  
states of 
$A^{3} \Sigma_{u}^{+}$,  $B^{3} \Pi_{g}$, $W^{3} \Delta_{u}$, 
${B'}^{3} \Sigma_{u}^{-}$, ${a'}^{1} \Sigma_{u}^{-}$, 
$a^{1} \Pi_{g}$, $w^{1} \Delta_{u}$ 
and $C^{3} \Pi_{u}$ states. 
Based on their differential cross section data, 
Johnson et al.\cite{2005JGRA..11011311J} 
derived integral cross sections (ICSs) 
for these electron impact excitations. 
In general, their ICSs are smaller than the 
other experimental cross sections at low impact energies below 
30 eV. These deviations may have some significance on 
study of atmospheric emissions, because a mean kinetic energy of 
electron at high altitudes is about 10 eV\cite{Wayne2000}. 
To shed light on this situation from a theoretical point of view, 
we perform the ab initio R-matrix calculations of 
electron impact excitations of N$_2$ molecule in this work. 

%previous experiments    + methods
% Ajello and Shemansky  : optical emission measurement
% Mason and Newell      : detection of metastable (a\Pi) molecules
% Poparic et al.        : crossed beam
% Zubek                 : optical emission excitation func.
% Zetner and Trajmar    : crossed beam / energy loss spectra ?
% Cartwright et al.     : crossed beam / energy loss spectra ?
% Khakoo et al.         : crossed beam / energy loss spectra ?
Many previous experimental measurements have been focused on 
excitation to a specific electronic state. 
For example, Ajello and Shemansky\cite{JGeophysResSpacePhys.90.9845} and 
Mason and Newell\cite{JPhysB.20.3913} 
measured ICSs for electron impact excitation to the ${a}^1 \Pi_g$ state, 
whereas Poparic et al.\cite{ChemPhys.240.283}, 
Zubek\cite{JPhysB.27.573} and 
Zubek and King\cite{JPhysB.27.2613} measured 
cross sections for the ${C}^3 \Pi_u$ state. 
In addition to these works, Zetner and Trajmar \cite{Zetner1987}
reported excitation cross sections to 
the ${A}^{3} \Sigma_{u}^{+}$,  $B^{3} \Pi_{g}$, 
$W^{3} \Delta_{u}$ and $a^{1} \Pi_{g}$ states. 
So far, comprehensive measurements of the excitation to
the 8 lowest electronic states are limited to three groups of 
Cartwright et al. \cite{PhysRevA.16.1013}, 
Brunger and Teubner \cite{PhysRevA.41.1413} and 
Khakoo et al.\cite{2005PhRvA..71f2703K}.  
%check this part. 
The measurements of Brunger and Teubner \cite{PhysRevA.41.1413} include 
excitation DCSs for the ${E}^{3} \Sigma_{g}^{+}$ and 
${a''}^{1} \Sigma_{g}^{+}$ states in addition to the 8 lowest excited states. 
The DCSs of Brunger and Teubner\cite{PhysRevA.41.1413} and 
Khakoo et al.\cite{2005PhRvA..71f2703K} were later converted 
to ICSs by Campbell et al.\cite{JPhysB.34.1185} and 
Johnson et al.\cite{2005JGRA..11011311J}, respectively. 
%how are they different?
%is there something more to discuss here?
Detailed reviews on electron N$_2$ collisions 
can be found in Itikawa\cite{JPhysChemRefData.35.31} 
and Brunger and Buckman\cite{Br02}. 

%previous calculations   + methods
% mention something on elastic scattering ??
Several groups have performed theoretical calculation of 
low energy electron collisions with N$_2$ molecule.  
For example, Chung and Lin\cite{PhysRevA.6.988} employed the Born 
approximation to calculate excitation cross sections for the 11 target states 
including the $A^{3} \Sigma_{u}^{+}$,  $B^{3} \Pi_{g}$, 
$W^{3} \Delta_{u}$, $a^{1} \Pi_{g}$, $w^{1} \Delta_{u}$ 
and $C^{3} \Pi_{u}$ states. 
Later, the same group of Holley et al.\cite{PhysRevA.24.2946}
calculated excitation ICSs for the $a^{1} \Pi_{g}$ state 
using a two-state-close-coupling method. 
Fliflet et al.\cite{JPhysB.12.3281} and 
Mu-Tao and McKoy\cite{PhysRevA.28.697} reported 
distorted-wave cross sections for excitation of 
the $A^{3} \Sigma_{u}^{+}$, $B^{3} \Pi_{g}$, 
$W^{3} \Delta_{u}$, $w^{1} \Delta_{u}$, $C^{3} \Pi_{u}$,  
$E^{3} \Sigma_{g}^{+}$, ${b'}^{1} \Sigma_{u}^{+}$ and 
${c'}^{1} \Sigma_{u}^{+}$ states. 
In general, these approximate methods are expected to 
be accurate at high impact energies above 30 eV.  
However, more elaborate method is required for precise comparison 
with experiment at low energies. 
Gillan et al.\cite{JPhysB.23.L407} calculated excitation ICSs for 
the $A^{3} \Sigma_{u}^{+}$, $B^{3} \Pi_{g}$ and 
$W^{3} \Delta_{u}$ states using the fixed nuclei R-matrix method. 
They included the 4 lowest target states in their R-matrix model, 
with target CI wave functions containing 2-13 CSFs.  
Their cross sections for the $A^{3} \Sigma_{u}^{+}$ and 
$W^{3} \Delta_{u}$ states agree well with the experimental results 
of Cartwright et al. \cite{PhysRevA.16.1013}. 
However, ICSs for the $B^{3} \Pi_{g}$ state 
deviate considerably from the experimental cross sections. 
Subsequently, they extended their R-matrix model to include 
the 8 lowest valence states\cite{JPhysB.29.1531}. 
Their target CI wave functions were much improved from 
their previous work by employing valence active space description, 
resulting in 68-120 CSFs per target state. 
In their paper, the ICSs were shown for the $A^{3} \Sigma_{u}^{+}$, 
$B^{3} \Pi_{g}$, $W^{3} \Delta_{u}$ and ${B'}^{3} \Sigma_{u}^{-}$ states,   
while the DCSs were presented for only the $A^{3} \Sigma_{u}^{+}$ state. 
Agreement with the ICSs of Cartwright et al.\cite{PhysRevA.16.1013} 
is good for these 4 excited states. 
However, agreement is marginal at DCS level. 

%about our/this work
% recent works on e-O2 excitation
%   --> relatively good agreement 
%   --> apply the procedure to the similar process of e-N2 excitation
%       <== make this statement first.
In this work, we study electron impact excitation of 
N$_2$ molecule by the fixed nuclei R-matrix method 
as in our previous work on electron O$_2$ 
scatterings \cite{PhysRevA.73.052707, 2006TASHIRO-2}. 
Although theoretical treatment is similar to the previous work of 
Gillan et al.\cite{JPhysB.29.1531}, more target states and 
partial waves of a scattering electron are included in the present work. 
Main purpose of this work is comparison of ICSs as well as DCSs 
for the 8 lowest excited states 
with the experimental results of Cartwright et al. \cite{PhysRevA.16.1013}, 
Brunger and Teubner\cite{PhysRevA.41.1413}, 
Campbell et al.\cite{JPhysB.34.1185}, Khakoo et al. \cite{2005PhRvA..71f2703K} 
and Johnson et al.\cite{2005JGRA..11011311J}. 
This is because previous theoretical works have covered only a part of 
these 8 excitations. 

%outline of this paper
In this paper, details of the calculation are presented in section 2, 
and we discuss the results in section 3 comparing our ICSs and DCSs with 
the previous theoretical and available experimental data.  
Then summary is given in section 4. 

\clearpage

\section{Theoretical methods}

\begin{figure}
\includegraphics{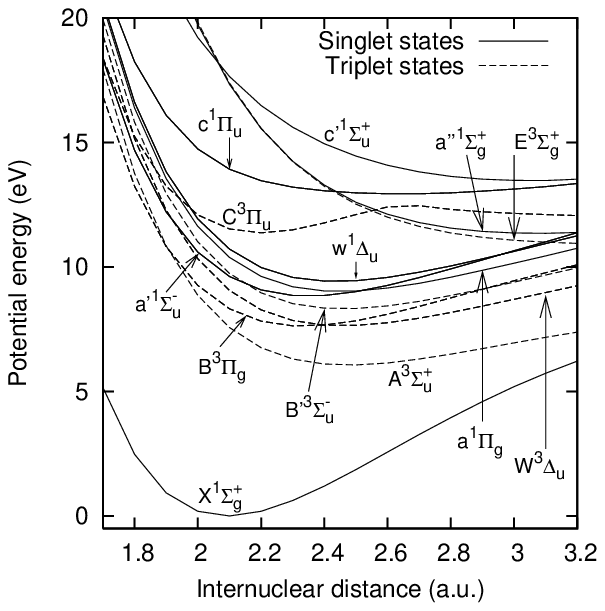}%
 \caption{\label{fig1} 
   Potential energy curves of the N$_2$ electronic states.
   The equilibrium distance of the ${X}^1 \Sigma^{+}_{g}$ state, 
   $R$ = 2.068 a$_0$ is used in our R-matrix calculations.  
   }
\end{figure}

The R-matrix method itself has been described extensively in the literature 
\cite{Bu05,Go05,Mo98} as well as 
in our previous paper\cite{PhysRevA.73.052707}. 
Thus we do not repeat general explanation of the method here. 
We used a modified version of the polyatomic programs in the UK molecular 
R-matrix codes \cite{Mo98}.  
These programs utilize the gaussian type orbitals (GTO) to 
represent target electronic states as well as a scattering electron. 
Although most of the previous R-matrix works in electron N$_2$ collisions 
have employed Slater type orbitals (STO), we select GTO mainly because of 
simplicity of the input and availability of basis functions. 
In the R-matrix calculations, we have included 13 target states; 
${X}^1 \Sigma^{+}_{g}$, $A^{3} \Sigma_{u}^{+}$, $B^{3} \Pi_{g}$, 
$W^{3} \Delta_{u}$, ${B'}^{3} \Sigma_{u}^{-}$, ${a'}^{1} \Sigma_{u}^{-}$, 
$a^{1} \Pi_{g}$, $w^{1} \Delta_{u}$, $C^{3} \Pi_{u}$, 
${E}^{3} \Sigma_{g}^{+}$, ${a''}^{1} \Sigma_{g}^{+}$, 
$c^{1} \Pi_{u}$ and ${c'}^{1} \Sigma_{u}^{+}$. 
The potential energy curves of these target electronic states are
shown in figure \ref{fig1} for reference. 
These target states were represented by valence configuration interaction 
wave functions constructed by state averaged complete active space SCF 
(SA-CASSCF) orbitals.  
Note that some target states, ${E}^{3} \Sigma_{g}^{+}$, 
${a''}^{1} \Sigma_{g}^{+}$ and ${c'}^{1} \Sigma_{u}^{+}$, 
are Rydberg states and cannot be described 
adequately in the present valence active space. 
Inclusion of these states are intended to improve quality of 
the R-matrix calculations by adding more target states in the 
model, as in our previous works \cite{PhysRevA.73.052707,2006TASHIRO-2} 
as well as other R-matrix works \cite{No92,Hi94}. 
Test calculation was performed with an extra $4 a_g$ orbital in the target 
orbital set. However, the target excitation energies as well as the 
excitation cross sections did not change much compared to the results 
with valence orbital set described above. 
Also, removal of $3 b_{1u}$ orbital from target active space 
did not affect the result much in our calculation. 
In this study, the SA-CASSCF orbitals were obtained by calculations with 
MOLPRO suites of programs \cite{molpro}. 
The target orbitals were constructed from the [5s3p1d] level of 
basis set taken from Sarpal et al. \cite{Sa96}. 
In our fixed-bond R-matrix calculations, the target states were 
evaluated at the equilibrium bond length $R$ = 2.068 a$_0$ 
of the N$_2$ ${X}^1\Sigma^{+}_{g}$ ground electronic state. 
%% state something ``we have performed two sets of calculations
%% with R=2.068 and 2.10 a0
%% Since there is no noticable difference, apart from slight change 
%% of resonance peak, we only present R=2.068a0 results in 
%% the discussion section.'' 
Although we also performed calculations with $R$ = 2.100 a$_0$ as in 
the previous R-matrix calculation of Gillan et al.\cite{JPhysB.29.1531}, 
the cross sections with $R$ = 2.068 a$_0$ and $R$ = 2.100 a$_0$ are 
almost the same. Thus, we will only show the results with 
the equilibrium bond length of N$_2$ in the next section. 
The radius of the R-matrix sphere was chosen to be 10 a$_0$ in our  
calculations.
In order to represent the scattering electron, we included diffuse
gaussian functions up to $l$ = 5, with 9 functions for $l$ = 0, 7 functions 
for $l$ = 1 - 3 and 6 functions for $l$ = 4 and 5.  
Exponents of these diffuse gaussians were fitted using the GTOBAS 
program \cite{Fa02} in the UK R-matrix codes.  
In addition to these continuum orbitals, we included 8 extra virtual 
orbitals, one for each symmetry. 

We constructed the 15-electron configurations from the orbitals
listed in table \ref{tab0}. 
The CI target wave functions are composed of the valence orbitals in 
table \ref{tab0} with the 1$a_g$ and 1$b_{1u}$ orbitals kept doubly 
occupied. 
We have included 3 types of configurations in the calculation. 
The first type of configurations has the form,  
\begin{equation}
1a_g^2 1b_{1u}^2 \{ 2a_g 3 a_g 1 b_{2u} 1 b_{3u} 2 b_{1u} 3 b_{1u} 
1 b_{3g} 1 b_{2g} \}^{10}  
\left( {}^{1} A_{g} \right) \{5a_{g}...39a_{g} \}^{1} 
\left( {}^{2}A_g \right), 
\end{equation}
here we assume that the total symmetry of this 15 electrons system is
${}^2A_g$. 
The first 4 electrons are always kept in the 1$a_g$ and 1$b_{1u}$
orbitals, then the next 10 electrons are distributed over the valence
orbitals with restriction of target state symmetry, ${}^{1} A_{g}$
symmetry of the N$_2$ ground state in this case. 
The last electron, the scattering electron, occupies one of the
diffuse orbitals, of $a_{g}$ symmetry in this example. 
To complete the wave function with the total symmetry ${}^2A_g$, 
we also have to include configurations with the other target states 
combined with diffuse orbitals having appropriate symmetry in the same
way as in the example. 
The second type of configurations has the form,  
\begin{equation} 
1a_g^2 1b_{1u}^2 \{ 2a_g 3 a_g 1 b_{2u} 1 b_{3u} 2 b_{1u} 3 b_{1u} 
1 b_{3g} 1 b_{2g} \}^{10}  
\left( {}^{1} A_{g} \right) \{ 4a_{g} \}^{1} \left( {}^{2}A_g
\right),
\end{equation}
where the scattering electron occupies a bound $4a_{g}$ extra virtual 
orbital, instead of the diffuse continuum orbitals in the 
expression (1). 
As in table \ref{tab0}, we included one extra virtual orbital for each
symmetry. 
The third type of configurations has the form, 
\begin{equation}
1a_g^2 1b_{1u}^2 \{ 2a_g 3 a_g 1 b_{2u} 1 b_{3u} 2 b_{1u} 3 b_{1u} 1
b_{3g} 1 b_{2g} \}^{11} 
\left( {}^{2}A_g \right).
\end{equation} 
In this case, the last 11 electrons including the scattering electron 
are distributed over the valence orbitals with the restriction of 
${}^2A_g$ symmetry. 
Note that the third type of configurations are crucial in
description of N$_2^-$ resonance states, which often have dominant
contributions to the excitation cross sections. 
In this way, the number of configurations generated for a specific total 
symmetry is typically about 60000, though the final dimension of the inner 
region Hamiltonian 
is reduced to be about 600 by using CI target contraction and 
prototype CI expansion method \cite{Te95}. 

The R-matrix calculations were performed for all 8 irreducible 
representations of the D$_{2h}$ symmetry, 
$A_g$, $B_{2u}$, $B_{3u}$, $B_{1g}$, $B_{1u}$, $B_{3g}$, $B_{2g}$ 
and $A_u$, in doublet spin multiplicity of the
electron plus target system. 
DCSs were evaluated in the same way as 
in our previous paper\cite{2006TASHIRO-2}. 

\clearpage

\section{Results and discussion}

\subsection{Excitation energies}

Figure \ref{fig1} shows the potential energy curves of all N$_2$ target 
states included in the present R-matrix model. 
These curves were obtained by the same SA-CASSCF method employed 
in our R-matrix calculation. 
Table \ref{tab1} compares the excitation energies of the N$_2$ target states 
from the present calculation with the previous R-matrix results of 
Gillan et al.\cite{JPhysB.29.1531}, multi-reference coupled cluster results of 
Ben-Shlomo and Kaldor \cite{JChemPhys.92.3680} as well as experimental values. 
Since these energies are evaluated at different inter-nuclear distance, 
2.068 $a_0$ in our case, 
2.100 $a_0$ in Gillan et al.\cite{JPhysB.29.1531} and 
2.074 $a_0$ in Ben-Shlomo and Kaldor \cite{JChemPhys.92.3680}, 
precise comparison is not so meaningful. 
However, deviations of excitation energies from the experimental 
values are less than 0.8 eV in our calculation, which 
is good considering the level of calculation. 
In terms of excitation energies, our calculation and 
the previous R-matrix calculation of Gillan et al.\cite{JPhysB.29.1531} 
have similar quality. 

In addition to this good agreement of target energies with experimental
results, N$_2^+$ energies are also well described in our SA-CASSCF
calculation. In our calculation, N$_2^+$ $X {}^2 \Sigma_g^+$ and 
$A {}^2 \Pi_u$ states are located 
at 15.63 and 17.21 eV above N$_2$ $X {}^1 \Sigma_g^+$ state, respectively. 
Compared to the experimental values of 15.61 and 17.08 eV, 
our SA-CASSCF calculation gives good results. 
Note that the energy ordering of N$_2^+$ $X {}^2 \Sigma_g^+$ and 
$A {}^2 \Pi_u$ states are not well described in 
the Hartree Fock level calculation, see Ermler and McLean \cite{JChemPhys.73.2297} for
example.

\subsection{Integral cross sections}

Figure \ref{fig2} shows integral cross sections for electron impact 
excitation from the N$_2$ $X^{1} \Sigma_{g}^{+}$ state to the 
$A^{3} \Sigma_{u}^{+}$,  $B^{3} \Pi_{g}$, $W^{3} \Delta_{u}$ 
and ${B'}^{3} \Sigma_{u}^{-}$ states. 
In this figure, present results are compared with 
the previous R-matrix calculations of Gillan et al.\cite{JPhysB.29.1531}, 
recent calculations of da Costa and Lima \cite{2006IJQC..106.2664D}, 
experimental results of Cartwright et al. \cite{PhysRevA.16.1013}, 
Campbell et al.\cite{JPhysB.34.1185} and recent measurements of 
Johnson et al.\cite{2005JGRA..11011311J}.  
Renormalized values of Cartwright et al. \cite{PhysRevA.16.1013} are 
used as recommended by Trajmar et al. \cite{PhysRep.97.221}. 
Figure \ref{fig3} compares the present excitation cross sections 
of the ${a'}^{1} \Sigma_{u}^{-}$, $a^{1} \Pi_{g}$, $w^{1} \Delta_{u}$ 
and $C^{3} \Pi_{u}$ states with the previous experimental results of 
Cartwright et al. \cite{PhysRevA.16.1013}, 
Campbell et al.\cite{JPhysB.34.1185} and 
Johnson et al.\cite{2005JGRA..11011311J}.
For the $a^{1} \Pi_{g}$ state cross sections, 
the recent calculations of 
da Costa and Lima \cite{2006IJQC..106.2664D},
other experimental values 
of Ajello and Shemansky \cite{JGeophysResSpacePhys.90.9845}, 
Zetner and Trajmar \cite{Zetner1987} and 
Mason and Newell \cite{JPhysB.20.3913} are included. 
For the $C^{3} \Pi_{u}$ state cross sections, 
the experimental results of Zubek \cite{JPhysB.27.573}, 
Zubek and King \cite{JPhysB.27.2613} and 
Poparic et al. \cite{ChemPhys.240.283} are included. 

Our excitation cross sections for the $A^{3} \Sigma_{u}^{+}$ state  
have a resonance feature at approximately 12 eV 
as in the previous R-matrix results of Gillan et al.\cite{JPhysB.29.1531}. 
The N$_2^-$ ${}^2 \Pi_u$ resonance state is responsible for 
this peak structure. 
The main configuration of this resonance state
is $1\pi_u^3 1\pi_g^2$. 
Other than the ${}^2 \Pi_u$ symmetry partial cross sections, 
The ${}^2 \Pi_g$ symmetry contributes to the ICSs as 
a smooth background component (not shown in the figure).  
Compared to the previous R-matrix cross sections, 
the peak at 12 eV is more pronounced in our case. 
Our results are slightly larger than theirs at 12-17.5 eV.  
Compared to the recent experimental results of 
Johnson et al.\cite{2005JGRA..11011311J}, 
our cross sections are about 50\% larger at 12.5-20 eV, though 
50\% smaller at 10 eV. Also our calculation overestimates 
the results of Campbell et al.\cite{JPhysB.34.1185}, 
however, the results of Cartwright et al. \cite{PhysRevA.16.1013} 
agree well with our results except at 12.5 eV. 
The position of the resonance peak depends rather strongly on the 
inter-nuclear distance of N$_2$ molecule, which is 12.2 eV 
for 2.068 $a_0$ and 11.75 eV for 2.100 $a_0$ in our calculations. 
Thus, inclusion of vibrational motion may be necessary to resolve 
this discrepancy of the resonance peak. 

Our excitation cross sections for the $B^{3} \Pi_{g}$ state 
have a small bump at 12.8 eV, which is not evident in 
the previous R-matrix cross sections. 
%The ${}^2 \Delta_g$ symmetry partial cross sections contribute 
%to this small bump. 
The origin of this bump is the N$_2^-$ ${1}^2 \Delta_g$ state, with main
configuration of $3\sigma_g^1 1\pi_g^2$.
%Add some remark on this ``resonance'' at the end of section II!
Other than this bump, the ICSs are mostly composed of the ${}^2 \Pi_g$ 
symmetry contribution and have a shape similar to 
the previous R-matrix results. 
The magnitude of our ICSs is about 50\% larger than the previous 
results of Gillan et al.\cite{JPhysB.29.1531}. 
Recently, da Costa and Lima \cite{2006IJQC..106.2664D} calculated 
ICSs for the $B {}^{3} \Pi_{g}$ state using the Schwinger multichannel
method with the minimal orbital basis for the single configuration 
interactions (MOB-SCI) approach. 
There cross sections are much larger than our results above 12 eV. 
Also, there is a prominent peak around 10 eV in their ICSs, which 
does not exist in the R-matrix calculations. 
Compared to the experimental ICSs, our results agree well with the cross
sections of Cartwright et al. \cite{PhysRevA.16.1013}, especially above 15 eV.  
However, the results of Campbell et al.\cite{JPhysB.34.1185} are 
much larger than ours. 
Recent measurements of Johnson et al.\cite{2005JGRA..11011311J} agree 
better with the previous R-matrix calculation of 
Gillan et al.\cite{JPhysB.29.1531}.   

For the excitation cross sections for the $W^{3} \Delta_{u}$ state, 
our results have a shape and magnitude similar to 
the previous R-matrix results. 
Most of our ICSs are composed of the ${}^2 \Pi_g$ symmetry 
partial cross sections. 
%% {\bf Write something about small bump at 13.4eV, as in case of B state excitation.}
Agreement with the experimental cross sections of 
Johnson et al.\cite{2005JGRA..11011311J} is good in this case. 
The cross sections of Campbell et al.\cite{JPhysB.34.1185} 
agree well with our results at 15 and 17.5 eV, 
but their value is about half as much as our result at 20 eV. 
The results of Cartwright et al. \cite{PhysRevA.16.1013} are 
about two times larger than our cross sections.  

Our excitation cross sections for the ${B'}^{3} \Sigma_{u}^{-}$ state 
are about half of the previous R-matrix cross sections of 
Gillan et al.\cite{JPhysB.29.1531}. 
%why?
Apart from this difference in magnitude, the shape of the 
cross sections is similar. 
Dominant component in these ICSs is the ${}^2 \Pi_g$ symmetry partial 
cross sections, 
although the ${}^2 \Pi_u$ symmetry also has certain contribution 
around 18-20 eV. 
Among 3 different experimental measurements, our results agree 
well with the results of Johnson et al.\cite{2005JGRA..11011311J}. 
The experimental cross sections of other two groups 
are much larger than our results at 15 and 17.5 eV, and have 
a different energy dependence compared to the present calculation. 

The situation of the excitation cross sections for 
the ${a'}^{1} \Sigma_{u}^{-}$ state is similar to the case 
of the ${B'}^{3} \Sigma_{u}^{-}$ state. 
The ${}^2 \Pi_g$ and ${}^2 \Pi_u$ symmetry partial cross sections 
contribute almost equally to the ICSs. 
Our cross sections roughly agree with the results of 
Johnson et al.\cite{2005JGRA..11011311J}, while the cross sections of 
Cartwright et al. \cite{PhysRevA.16.1013} and 
Campbell et al.\cite{JPhysB.34.1185} at 15 eV are much larger than our result. 
The results of Cartwright et al. \cite{PhysRevA.16.1013} and 
Campbell et al.\cite{JPhysB.34.1185} 
decrease as impact energy increases from 15 to 20 eV, however,  
our cross sections increase mildly in this energy region. 

%% see johnson et al. 
%% for extensive discution on different experimental data sets,
In case of excitation to the $a^{1} \Pi_{g}$ state, 
several other experimental results are available in addition 
to the measurements of Cartwright et al. \cite{PhysRevA.16.1013}, 
Campbell et al.\cite{JPhysB.34.1185}, 
and Johnson et al.\cite{2005JGRA..11011311J}.  
The cross section profiles of Johnson et al.\cite{2005JGRA..11011311J}, 
Ajello and Shemansky \cite{JGeophysResSpacePhys.90.9845}, 
Cartwright et al. \cite{PhysRevA.16.1013} and 
Mason and Newell \cite{JPhysB.20.3913} are similar to our ICSs. 
However, the magnitude of our cross sections is 
lower than the experimental values in most case except 
the cross sections of Johnson et al.\cite{2005JGRA..11011311J}.  
At 15, 17.5 and 20 eV, agreement of our results with 
the cross sections of Johnson et al.\cite{2005JGRA..11011311J} 
is very good, although our cross section at 12.5 eV is 
twice as large as their value. 
Note that there is no dominant symmetry contribution to the calculated 
ICSs. All partial cross sections contribute rather equally to the ICSs. 
Recent ICSs of da Costa and Lima \cite{2006IJQC..106.2664D} by 
the Schwinger multichannel method are also shown in the panel (b) of the
figure \ref{fig3}. Their result has a sharp peak at 12 eV as in their 
calculation for the $B {}^{3} \Pi_{g}$ state excitation. 
This difference between our and their results may come from 
different number of target states considered in the scattering calculation. 
Only the $X {}^{1} \Sigma_{g}^1$, $a {}^{1} \Pi_{g}$ and $B {}^{3}
\Pi_{g}$ states were included in the calculations of da Costa and Lima.    
The other part of the cross section profile is similar to the
shape of our cross sections, although the magnitude of their cross
sections are about twice as large as our results at 15-20 eV.

Our excitation cross section for the $w^{1} \Delta_{u}$ state 
gradually increases as a function of energy from the threshold to 
the broad peak around 17.5 eV, then decreases toward 20 eV. 
In this case, agreement with the results of 
Johnson et al.\cite{2005JGRA..11011311J} is not so good compared 
to the excitations of the $a^{1} \Pi_{g}$ and 
${a'}^{1} \Sigma_{u}^{-}$ states. 
Our cross sections are about 50\% larger than their values at 17.5 and 20 eV. 
At 15 eV, our results agree well with the cross section of 
Johnson et al.\cite{2005JGRA..11011311J}, however, they are 
about 50\% lower than the results of 
Cartwright et al. \cite{PhysRevA.16.1013} and 
Campbell et al.\cite{JPhysB.34.1185}.  
In the calculated ICSs, the ${}^2 \Pi_u$ symmetry partial cross section is 
a major component, with a minor contribution from the ${}^2 \Pi_g$ symmetry.

The calculated excitation cross sections for 
the $C^{3} \Pi_{u}$ state has a peak  
similar to the experimental results of Zubek \cite{JPhysB.27.573} and 
Poparic et al.\cite{ChemPhys.240.283}.   
Although the shape of the cross sections is similar, 
position of the cross section peak is different from 
experimental results.  
In our case, it is located at about 
17 eV, whereas corresponding peaks are located at 
14 eV in the experimental cross sections. 
The height of the peak in our ICSs is lower than the 
experimental values of Zubek \cite{JPhysB.27.573} 
and Poparic et al.\cite{ChemPhys.240.283}.  
It is unclear whether there is a cross section peak in  
the experimental cross sections of Cartwright et al. \cite{PhysRevA.16.1013}, 
Campbell et al.\cite{JPhysB.34.1185} 
and Johnson et al.\cite{2005JGRA..11011311J}.  
At least, it appears that they do not have a peak around 17 eV. 
The origin of this discrepancy in the cross section peak is uncertain, 
but may be related to the employment of the fixed-nuclei approximation or 
insufficiency of higher excited target states in the R-matrix model. 
The calculated ICSs are composed of the ${}^2 \Sigma_u^+$ and 
${}^2 \Sigma_u^-$ symmetry partial cross sections near the peak 
structure at 17 eV. 
The contribution of the ${}^2 \Sigma_u^+$ symmetry is about 50\% larger 
than the ${}^2 \Sigma_u^-$ component. 
Other than these two symmetries, the ${}^2 \Pi_g$ symmetry partial cross 
section contributes to the ICSs as a smooth background component. 
%
%why does this deviation occures? 
%

\subsection{Differential cross sections}

\begin{figure}
\includegraphics{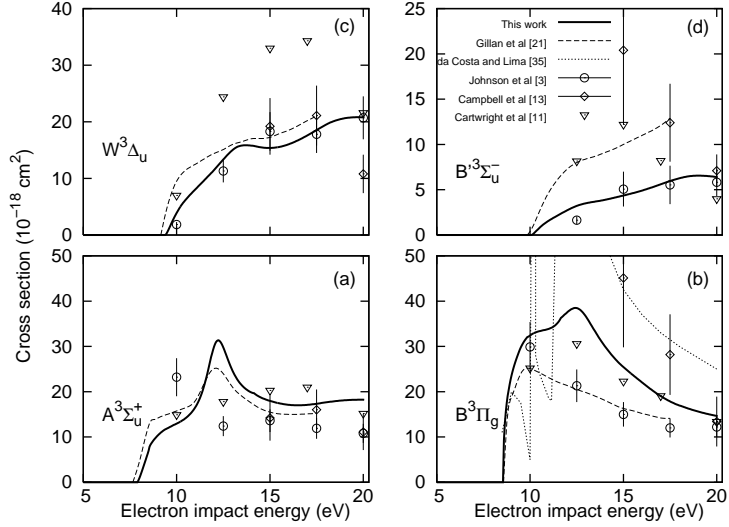}%
 \caption{\label{fig2} 
  Integral excitation cross sections of the 
  $A^{3} \Sigma_{u}^{+}$ (panel a), 
  $B^{3} \Pi_{g}$ (panel b), $W^{3} \Delta_{u}$ (panel c) 
  and ${B'}^{3} \Sigma_{u}^{-}$ (panel d) states. 
  Our results are shown in thick full lines.  
  For comparison, we include the previous R-matrix results of 
  Gillan et al. \cite{JPhysB.29.1531}, Schwinger multichannel results 
  of da Costa and Lima \cite{2006IJQC..106.2664D},  
  the experimental cross sections of 
  Cartwright et al. \cite{PhysRevA.16.1013}, 
  Campbell et al.\cite{JPhysB.34.1185} and 
  Johnson et al.\cite{2005JGRA..11011311J}. 
   }
\end{figure}

\begin{figure}
 \includegraphics{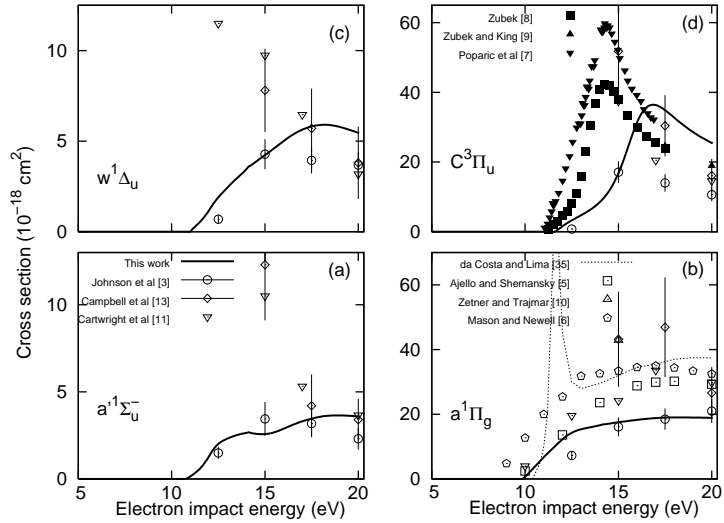}%
 \caption{\label{fig3} 
  Integral excitation cross sections of the 
  ${a'}^{1} \Sigma_{u}^{-}$ (panel a), 
  $a^{1} \Pi_{g}$ (panel b), $w^{1} \Delta_{u}$ (panel c) 
  and $C^{3} \Pi_{u}$ (panel d) states.
  Our results are shown in thick full lines. 
  In addition to the experimental ICSs in Fig.\ref{fig2}, 
  we include the results of 
  Ajello and Shemansky \cite{JGeophysResSpacePhys.90.9845}, 
  Zetner and Trajmar \cite{Zetner1987}, 
  Mason and Newell \cite{JPhysB.20.3913}, 
  Poparic et al. \cite{ChemPhys.240.283}, 
  Zubek \cite{JPhysB.27.573} and 
  Zubek and King \cite{JPhysB.27.2613}.  
  }
\end{figure}

\begin{figure}
 \includegraphics{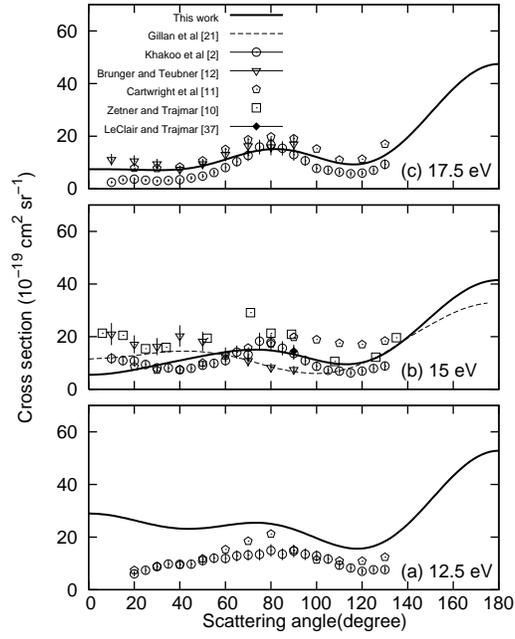}%
 \caption{\label{fig4} 
  Differential cross sections for electron impact excitation from the 
  N$_2$ ${X}^1\Sigma^{+}_{g}$ state to the ${A}^{3} \Sigma_u^+$ state. 
  Panel (a): electron impact energy of 12.5 eV, (b):15 eV and (c): 17.5 eV.   
  Full line represents our result. 
  For comparison, we include the previous theoretical  
  cross sections of Gillan et al.\cite{JPhysB.29.1531}, 
  experimental results of Khakoo et al. \cite{2005PhRvA..71f2703K},  
  Brunger and Teubner \cite{PhysRevA.41.1413}, 
  Cartwright et al. \cite{PhysRevA.16.1013}, 
  Zetner and Trajmar \cite{Zetner1987} and 
  LeClair and Trajmar \cite{JPhysB.29.5543}. 
   }
\end{figure}

Figure \ref{fig4} shows calculated DCSs for excitation of 
the ${A}^{3} \Sigma_{u}^{+}$ state with the experimental 
results of Khakoo et al. \cite{2005PhRvA..71f2703K}, 
Brunger and Teubner \cite{PhysRevA.41.1413}, 
Cartwright et al. \cite{PhysRevA.16.1013}, 
Zetner and Trajmar \cite{Zetner1987}, 
LeClair and Trajmar \cite{JPhysB.29.5543} and 
the previous R-matrix DCSs of Gillan et al.\cite{JPhysB.29.1531}.   
Our DCSs at 12.5, 15 and 17.5 eV have similar shape in common.  
They are enhanced in backward direction and have a small dimple 
at 120 degrees with a bump at 75 degrees. 
At 17.5 eV, our cross sections are located between the 
experimental values of Khakoo et al.\cite{2005PhRvA..71f2703K} 
and Cartwright et al. \cite{PhysRevA.16.1013}.   
The profile of the experimental DCSs are reproduced well in 
our calculation. 
At 15 eV, our results agree better with the results of 
Khakoo et al. \cite{2005PhRvA..71f2703K} compared to the other experiments. 
In the DCSs of the previous R-matrix calculation of 
Gillan et al.\cite{JPhysB.29.1531},  
a bump is located at 40 degrees and a small dimple 
is located at 100 degrees, which agree better with the 
experimental results of Brunger and Teubner\cite{PhysRevA.41.1413}. 
In our calculation, these dimple and bump are shifted toward 
backward direction by 20 degrees, and agreement with 
the results of Brunger and Teubner \cite{PhysRevA.41.1413} is not so good. 
At 12.5 eV, our calculation overestimates the experimental results by  
a factor of two. As seen in panel (a) of Fig.\ref{fig2}, 
this discrepancy is related to the existence of a resonance 
peak around 12.5 eV. 

\begin{figure}
 \includegraphics{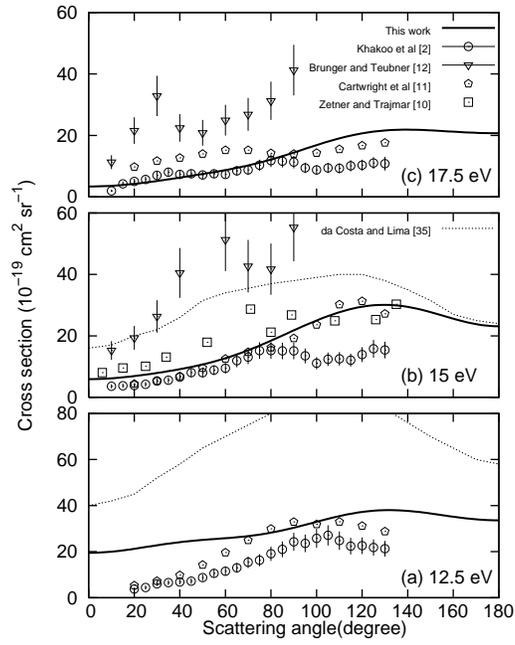}%
 \caption{\label{fig5}  
  Differential cross sections for electron impact excitation from the 
  N$_2$ ${X}^1\Sigma^{+}_{g}$ state to the ${B}^{3} \Pi_g$ state. 
  Panel (a): electron impact energy of 12.5 eV, (b):15 eV and (c): 17.5 eV.   
  The results of Schwinger multichannel calculation by 
  da Costa and Lima \cite{2006IJQC..106.2664D} are also shown in the panels.    
  Other details are the same as in Fig.\ref{fig4}. 
   }
\end{figure}

Figure \ref{fig5} compares calculated excitation DCSs for the 
$B^{3} \Pi_{g}$ state with the experimental and recent theoretical results. 
Our DCSs at 12.5, 15 and 17.5 eV have backward-enhanced feature 
with a broad peak at 130 degrees. 
At 15 and 17.5 eV, our DCSs agree well with the results of 
Khakoo et al. \cite{2005PhRvA..71f2703K} 
at forward direction below 80 degrees. 
However, their DCSs are smaller than ours by a factor of 
two at 80-130 degrees. Agreement with the results of 
Cartwright et al. \cite{PhysRevA.16.1013} 
at 15 eV is good at 20-130 degrees, although their DCSs are twice as 
large as our DCSs at 17.5 eV for low scattering angles. 
Because of a resonance-like feature at 12.5 eV as seen in panel (b) 
of Fig.\ref{fig2}, our results are larger than the experimental results 
at 12.5 eV. 
Recent Schwinger multi-channel results of da Costa and Lima 
\cite{2006IJQC..106.2664D} are much larger than our DCSs at 12.5 
and 15 eV. The deviation is especially large at 12.5 eV, which is
possibly related to the difference in the excitation energies of 
the target state.

\begin{figure}
 \includegraphics{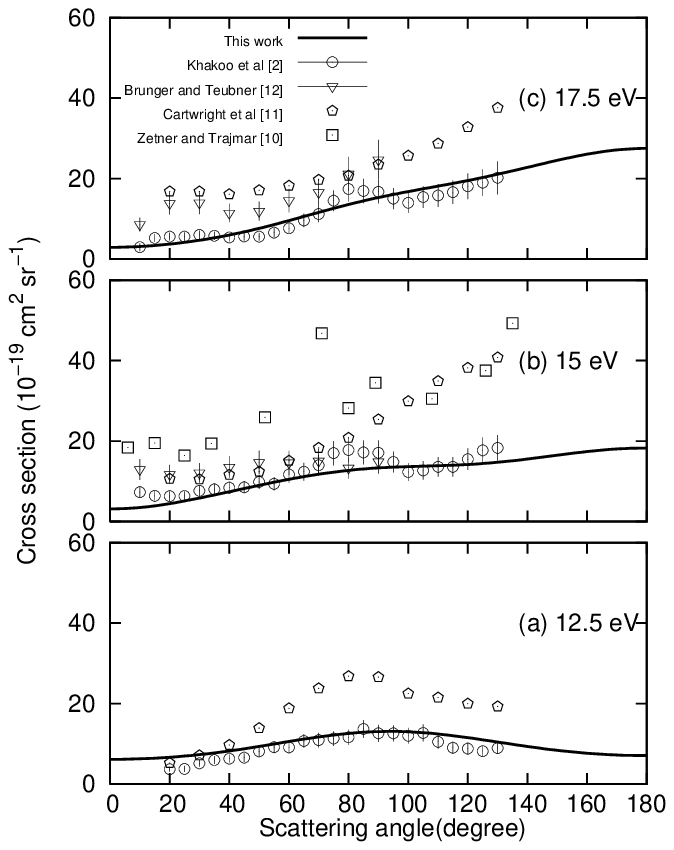}%
 \caption{\label{fig6} 
  Differential cross sections for electron impact excitation from the 
  N$_2$ ${X}^1\Sigma^{+}_{g}$ state to the ${W}^{3} \Delta_u$ state. 
  Panel (a): electron impact energy of 12.5 eV, (b):15 eV and (c): 17.5 eV.   
  Other details are the same as in Fig.\ref{fig4}. 
   }
\end{figure}

Figure \ref{fig6} shows the excitation DCSs for 
the $W^{3} \Delta_{u}$ state with the experimental cross sections. 
At 15 and 17.5 eV, our cross section gradually increases 
as a function of scattering angle, without noticeable bump or dip. 
At 12.5 eV, the shape of DCSs is nearly symmetric around 90 degrees. 
Agreement with the experimental DCSs of 
Khakoo et al. \cite{2005PhRvA..71f2703K} is good, 
although their results at 15 and 17.5 eV have more complex structure 
such as a small peak at 80 degrees. 
Our DCSs are generally smaller than the other experimental results 
of Brunger and Teubner\cite{PhysRevA.41.1413}, 
Cartwright et al. \cite{PhysRevA.16.1013}, 
Zetner and Trajmar \cite{Zetner1987}. 

\begin{figure}
 \includegraphics{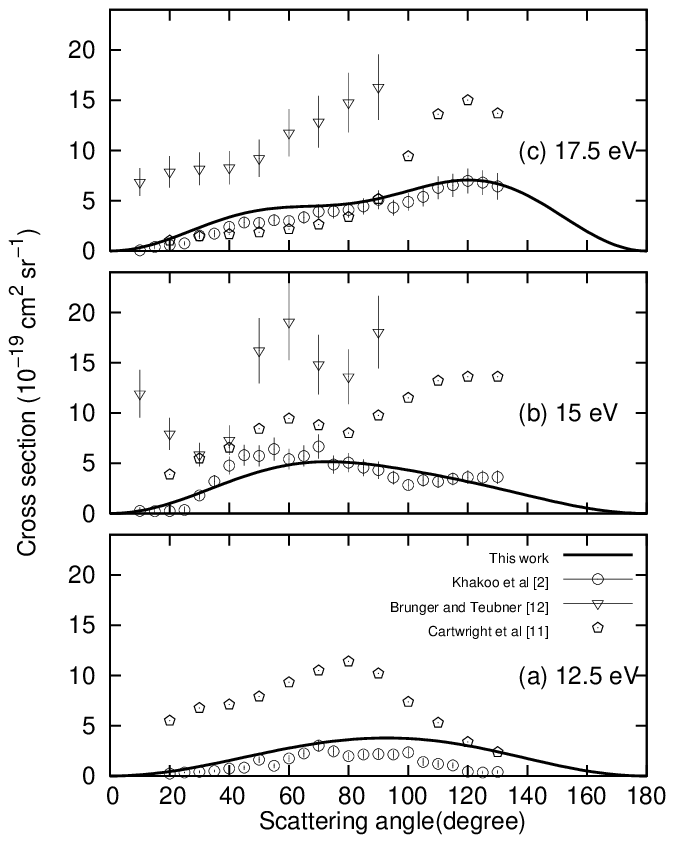}%
 \caption{\label{fig7}
  Differential cross sections for electron impact excitation from the 
  N$_2$ ${X}^1\Sigma^{+}_{g}$ state to the ${B'}^{3} \Sigma_u^-$ state. 
  Panel (a): electron impact energy of 12.5 eV, (b):15 eV and (c): 17.5 eV.   
  Other details are the same as in Fig.\ref{fig4}. 
   }
\end{figure}

Excitation cross sections for the ${B'}^{3} \Sigma_{u}^{-}$ state 
are shown in figure \ref{fig7}. 
Calculated DCSs decrease to be zero toward 0 and 180 degrees, 
because of a selection rule associated with 
$\Sigma^{+}$-$\Sigma^{-}$ transition \cite{Go71,Ca71}. 
%why 2 peaks at 17.5 eV and a single peak at 12.5 and 15 eV?
Our DCSs have a broad single peak near 90 degrees at 12.5 and 15 eV, 
whereas there are two broad peaks at 17.5 eV. 
The position of the right peak at 17.5 eV coincides with that of 
the experimental DCSs of Khakoo et al. \cite{2005PhRvA..71f2703K} and 
Cartwright et al. \cite{PhysRevA.16.1013}, 
although the peak of Cartwright et al. \cite{PhysRevA.16.1013} is 
much higher than ours. 
Our results agree well with the DCSs of 
Khakoo et al. \cite{2005PhRvA..71f2703K} at 15 and 17.5 eV. 
However, their cross sections at 15 eV have a small dip at 100 degrees 
and a small bump 60 degrees, which do not exist in our results. 
At 12.5 eV, our cross sections are slightly larger than the results 
of Khakoo et al.\cite{2005PhRvA..71f2703K}. 
On the whole, agreement with the other experimental results of 
Brunger and Teubner \cite{PhysRevA.41.1413} and 
Cartwright et al. \cite{PhysRevA.16.1013} is not good. 

\begin{figure}
 \includegraphics{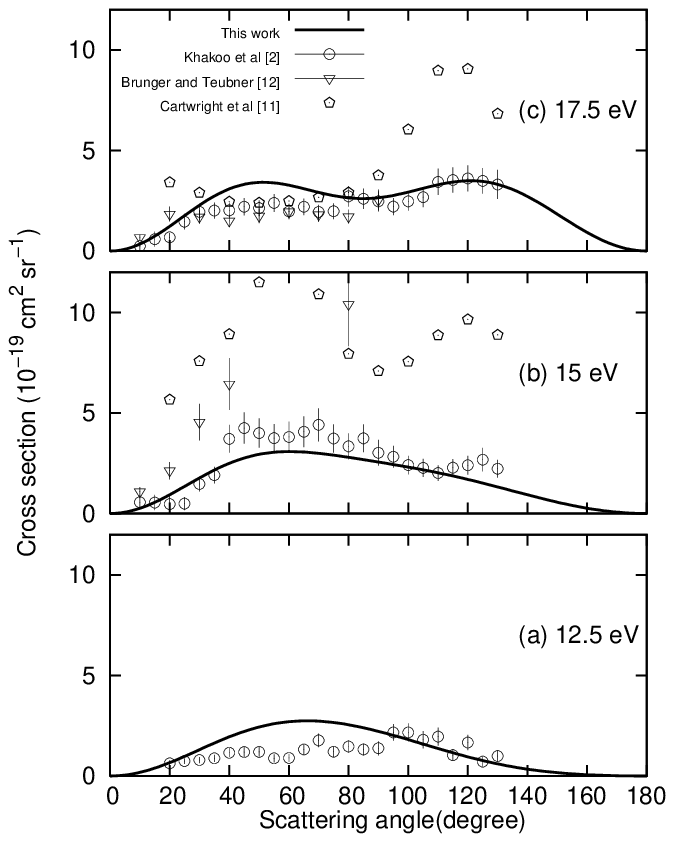}%
 \caption{\label{fig8} 
  Differential cross sections for electron impact excitation from the 
  N$_2$ ${X}^1\Sigma^{+}_{g}$ state to the ${a'}^{1} \Sigma_u^-$ state. 
  Panel (a): electron impact energy of 12.5 eV, (b):15 eV and (c): 17.5 eV.   
  Other details are the same as in Fig.\ref{fig4}. 
   }
\end{figure}

Figure \ref{fig8} shows the excitation DCSs for the 
${a'}^{1} \Sigma_{u}^{-}$ state. 
Because of $\Sigma^{+}$-$\Sigma^{-}$ selection rule, 
DCSs at 0 and 180 degrees become zero as in the case of 
the ${B'}^{3} \Sigma_{u}^{-}$ state DCSs. 
Calculated DCSs have a broad single peak near 60 degrees at 12.5 and 15 eV. 
At 17.5 eV, there are two broad peaks at 50 and 120 degrees. 
Although there is slight overestimation of DCSs near 50-60 degrees, 
our DCSs agree marginally with the results of 
Khakoo et al.\cite{2005PhRvA..71f2703K}. 
Agreement with the other experimental results is 
not good except low scattering angles at 17.5 eV.  

\begin{figure}
 \includegraphics{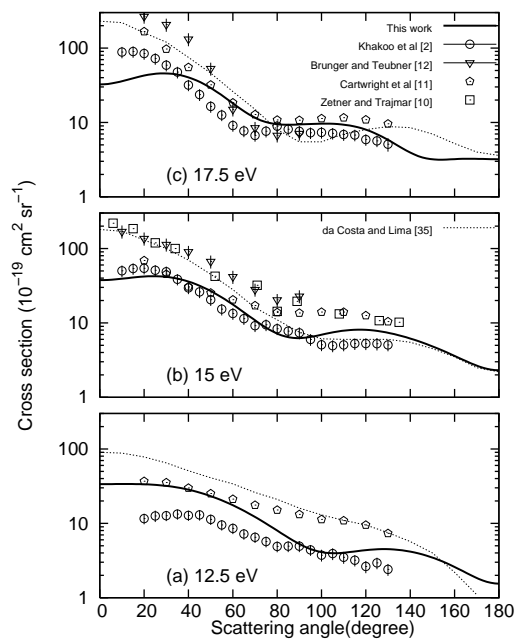}%
 \caption{\label{fig9} 
  Differential cross sections for electron impact excitation from the 
  N$_2$ ${X}^1\Sigma^{+}_{g}$ state to the ${a}^{1} \Pi_g$ state. 
  Panel (a): electron impact energy of 12.5 eV, (b):15 eV and (c): 17.5 eV.   
  Note that the DCSs are shown in logarithmic scale. 
  Other details are the same as in Fig.\ref{fig4}. 
   }
\end{figure}

Figure \ref{fig9} compares our excitation DCSs for 
the $a^{1} \Pi_{g}$ state with the experimental cross sections. 
Because of large variation of the DCSs, the cross sections are 
shown in logarithmic scale. 
Calculated DCSs are strongly forward-enhanced, 
which is consistent with all experimental results shown 
in the figure. 
Our DCSs at 12.5 eV have a small dip around 100 degrees, which 
moves forward to 85 degrees at 15 eV and 75 degrees at 17.5 eV. 
This behavior roughly agrees with the results of 
Cartwright et al.\cite{PhysRevA.16.1013} and 
Khakoo et al.\cite{2005PhRvA..71f2703K}. 
At 15 eV, our DCSs agree better with 
the results of Khakoo et al.\cite{2005PhRvA..71f2703K} than 
the other experimental DCSs. 
At 17.5 eV, the results of Cartwright et al.\cite{PhysRevA.16.1013}  
are closer to our DCSs at scattering angles above 40 degrees.  
Below 40 degrees, our calculation significantly underestimates 
the experimental DCSs. 
Our results at 12.5 eV are located between the DCSs of 
Cartwright et al. \cite{PhysRevA.16.1013} 
and Khakoo et al.\cite{2005PhRvA..71f2703K}, 
however the shape of the DCSs is similar to their results. 
The shapes of DCSs calculated by da Costa and Lima 
\cite{2006IJQC..106.2664D} are similar to our results. 
However, their cross sections are larger than our results at
low-scattering angles below 80 degrees, where their results agree better 
with the experimental DCSs of Brunger and Teubner \cite{PhysRevA.41.1413} and 
Zetner and Trajmar \cite{Zetner1987}.

\begin{figure}
 \includegraphics{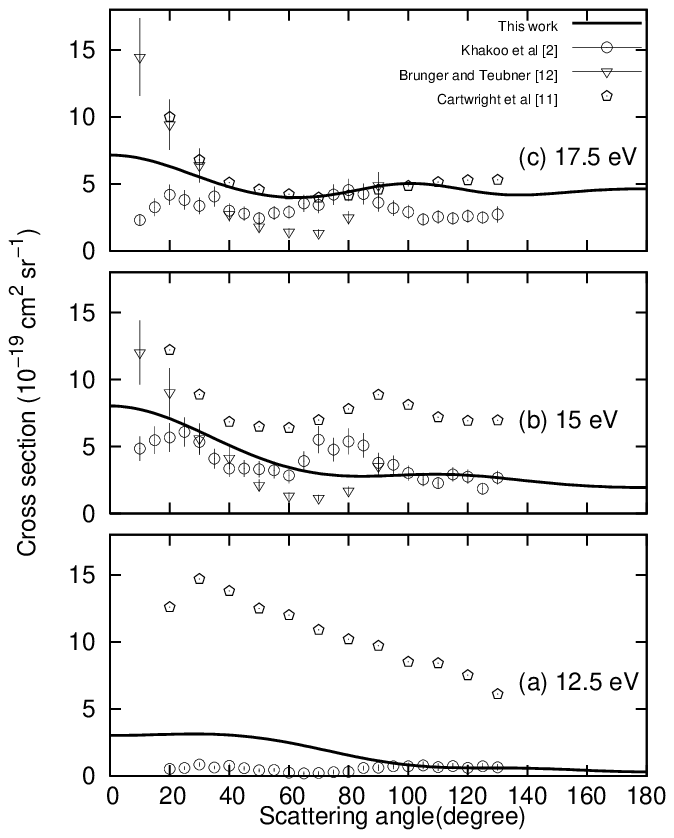}%
 \caption{\label{fig10} 
  Differential cross sections for electron impact excitation from the 
  N$_2$ ${X}^1\Sigma^{+}_{g}$ state to the ${w}^{1} \Delta_u$ state. 
  Panel (a): electron impact energy of 12.5 eV, (b):15 eV and (c): 17.5 eV.   
  Other details are the same as in Fig.\ref{fig4}. 
   }
\end{figure}

Figure \ref{fig10} shows calculated excitation DCSs for 
the $w^{1} \Delta_{u}$ state with the experimental cross sections. 
Our DCSs are enhanced in forward direction as in the case of 
the $a^{1} \Pi_{g}$ state. However, magnitude of the 
enhancement is much smaller than that of the $a^{1} \Pi_{g}$ state. 
Agreement with the DCSs of Cartwright et al. \cite{PhysRevA.16.1013} 
is good at 17.5 eV except low scattering angles below 20 degrees. 
At 12.5 and 15 eV, their results are much larger than our DCSs. 
At 15 eV, our DCSs agree marginally with the results of 
Khakoo et al. \cite{2005PhRvA..71f2703K}, 
although details of the DCS profile are different. 
Their results are smaller than ours at 17.5 and 12.5 eV. 
Discrepancy is especially large for forward scattering at 12.5 eV. 

\begin{figure}
 \includegraphics{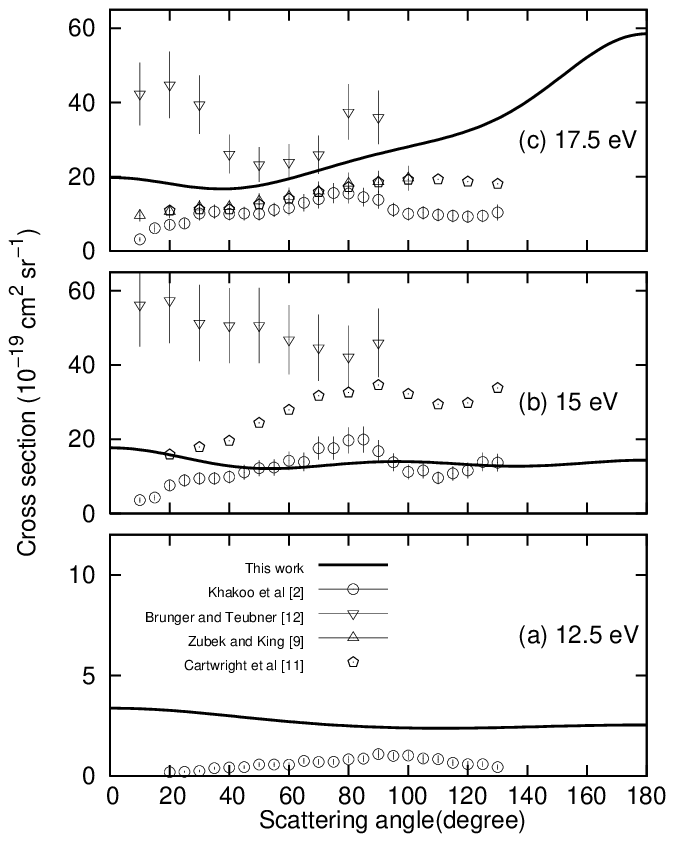}%
 \caption{\label{fig11} 
  Differential cross sections for electron impact excitation from the 
  N$_2$ ${X}^1\Sigma^{+}_{g}$ state to the ${C}^{3} \Pi_u$ state. 
  Panel (a): electron impact energy of 12.5 eV, (b):15 eV and (c): 17.5 eV.   
  The experimental DCSs of Zubek and King \cite{JPhysB.27.2613} are added.  
  Other details are the same as in Fig.\ref{fig4}. 
   }
\end{figure}

Figure \ref{fig11} shows excitation DCSs for the $C^{3} \Pi_{u}$ state 
with the experimental cross sections of 
Khakoo et al.\cite{2005PhRvA..71f2703K}, 
Brunger and Teubner\cite{PhysRevA.41.1413}, 
Zubek and King \cite{JPhysB.27.2613} and 
Cartwright et al. \cite{PhysRevA.16.1013}.  
Calculated DCS profiles are almost flat at 12.5 and 15 eV, 
whereas they are enhanced in backward direction at 17.5 eV. 
Below 90 degrees, slope of the calculated DCSs at 17.5 eV 
is similar to the results of Khakoo et al.\cite{2005PhRvA..71f2703K}, 
Zubek and King \cite{JPhysB.27.2613} and 
Cartwright et al. \cite{PhysRevA.16.1013}, 
though our results are about 50\% larger than their DCSs. 
In general, our results do not agree well with the experimental DCSs. 
Although the ICS of 
Khakoo et al. \cite{2005PhRvA..71f2703K} at 15 eV agrees well 
our result as shown in panel (d) of Fig.\ref{fig3}, 
angular dependence of the cross sections appears to be different. 
%analyze why?

%discussion on singlet-triplet forward/backward peak relation
%effect of bond length (2.07 <-> 2.10)

\subsection{Discussion}

%%
%%discussion concerning referee's comments
%%

% resonance-like peaks in Fig.2

The excitation ICSs of the $B^3 \Pi_g$ state, shown in panel (b) of figure
\ref{fig2}, have a small bump around 13 eV. However, there is no such
structure in the previous R-matrix ICSs of 
Gillan et al. \cite{JPhysB.29.1531}. 
The origin of this bump in our calculation is the N$_2^-$ ${1}^2
\Delta_g$ state, with main configuration of $3\sigma_g^1 1\pi_g^2$. 
The existence of the N$_2^-$ ${1}^2 \Delta_g$ state can also be verified by 
usual CASSCF calculation on N$_2^-$ with valence active space ignoring 
continuum orbitals. In molpro calculation, the energy of the 
${1}^2 \Delta_g$ state is 15.7 eV. Since diffuse continuum orbitals are
added in the R-matrix calculation, the energy of the state is stabilized
to be 12.8 eV in the present scattering calculation. 
In the same way, the N$_2^-$ ${}^2 \Pi_u$ ($1\pi_u^3 1\pi_g^2$) 
resonance peak in the $A{}^{3} \Sigma_u^+$ excitation ICSs can be 
verified by usual CASSCF calculation. In molpro calculation, 
it is located at 14.7 eV, whereas the position of the resonance 
is stabilized to be 12.2 eV in our R-matrix scattering calculation.  
%Also, a very small lump around 13 eV in the ICSs of the $W{}^{3} \Delta_u$ state 
%is related to the second lowest state of the N$_2^-$ ${}^2 \Pi_u$ 
%x($1\pi_u^3 1\pi_g^2$) state  
It is unclear why the bump in the ICSs of the $B^3 \Pi_g$ state  
is not evident in the previous R-matrix 
cross sections of Gillan et al.\cite{JPhysB.29.1531}.  
Some details of the R-matrix calculations are different in their
calculation and ours, e.g., they used hybrid orbitals with Slater type 
functions, whereas we employed SA-CASSCF orbitals with Gaussian type functions. 
These difference may contribute to the difference in magnitude of 
the ${}^2 \Delta_g$ partial cross section.

% Is FN OK?
In this study, we employed the fixed-nuclei (FN) approximation. 
As we can see in figure \ref{fig1}, equilibrium bond lengths of 
the excited N$_2$ states are longer than that of the ground state. 
%This situation is similar to the O$_2$ case.
Thus, in principle, it would be desirable to include the effect of 
nuclear motion in the R-matrix calculation. 
Use of the FN approximation may be responsible for 
several discrepancies between our calculation and experiments, 
including bumps in the ICSs of the $A^{3} \Sigma_{u}^{+}$ and 
$B^{3} \Pi_{g}$ states, the position 
of the peak in the ICSs of the $C^{3} \Pi_{u}$ state. 
Although the calculated DCSs agree very well with experimental results 
in general, our DCSs of the $A^{3} \Sigma_{u}^{+}$, $B^{3} \Pi_{g}$, 
$w^{1} \Delta_{u}$ and $C^{3} \Pi_{u}$ states at 12.5 eV are 2-4 times
larger than experimental results. These deviations in the near-threshold
DCSs can also be related to the FN approximation. 
In spite of these discrepancies, good agreements are observed between 
our calculation and experiments in most ICS and DCS cases as we can see 
in the figures. Agreements with the recent experimental results of 
Khakoo et al.\cite{2005PhRvA..71f2703K} and Johnson et al.
\cite{2005JGRA..11011311J} are especially impressive. 
It is possible to include nuclear motion in the R-matrix 
formalism through vibrational averaging of T-matrix elements 
or the non-adiabatic R-matrix method, though application of 
these methods will be a difficult task in the presence of many 
target electronic states. In the future, we plan to perform 
the R-matrix calculation with these methods including nuclear 
motion.

% effect of CI conf. / pol-cor terms on results

%other points / addition of further polirization effect..etc.

\clearpage

\section{summary}

We have investigated electron impact excitations of 
N$_2$ molecule using the fixed-bond R-matrix method 
which includes 13 target electronic states,
${X}^1 \Sigma^{+}_{g}$, $A^{3} \Sigma_{u}^{+}$, $B^{3} \Pi_{g}$, 
$W^{3} \Delta_{u}$, ${B'}^{3} \Sigma_{u}^{-}$, ${a'}^{1} \Sigma_{u}^{-}$, 
$a^{1} \Pi_{g}$, $w^{1} \Delta_{u}$, $C^{3} \Pi_{u}$, 
${E}^{3} \Sigma_{g}^{+}$, ${a''}^{1} \Sigma_{g}^{+}$, 
$c^{1} \Pi_{u}$ and ${c'}^{1} \Sigma_{u}^{+}$. 
These target states are described by CI wave functions in the valence 
CAS space, using SA-CASSCF orbitals. Gaussian type orbitals 
were used in this work, in contrast to the STOs in 
the previous R-matrix works. 
We have obtained integral cross sections as well as 
differential cross sections of excitations to the 
$A^{3} \Sigma_{u}^{+}$,  $B^{3} \Pi_{g}$, $W^{3} \Delta_{u}$, 
${B'}^{3} \Sigma_{u}^{-}$, ${a'}^{1} \Sigma_{u}^{-}$, 
$a^{1} \Pi_{g}$, $w^{1} \Delta_{u}$ 
and $C^{3} \Pi_{u}$ states, which have been studied a lot experimentally 
but not enough theoretically before. 
In general, good agreements are observed both in the 
integrated and differential cross sections, 
which is encouraging for further theoretical and experimental 
studies in this field.   
However, some discrepancies are seen in the integrated 
cross sections of the $A^{3} \Sigma_{u}^{+}$ 
and $C^{3} \Pi_{u}$ states, especially around a peak structure. 
Also, our DCSs do not agree well with the experimental results at 
low impact energy of 12.5 eV, compared to the higher energies 
of 15 and 17.5 eV. 
These discrepancies may be related to the fixed-nuclei approximation or 
insufficiency of higher excited target states in the R-matrix model.

%%%%%%%%%%%%%%%%%%%%%%%%%%%%%%%%%%%%%%%%%%%%%%%%%%%%%%%%%%%%%%%%%%%%%%%%%%

\begin{acknowledgments}
The work of M.T. is supported by the Japan Society for the 
Promotion of Science Postdoctoral Fellowships for Research Abroad. 
The present research is supported in part by the grant from the Air Force 
Office of Scientific Research: the Advanced High-Energy Closed-Cycle Chemical 
Lasers project (PI: Wayne C. Solomon, University of Illinois, 
F49620-02-1-0357). 
\end{acknowledgments}

\clearpage

%\bibliography{draft}
%%draft.bbl

\clearpage

% If in two-column mode, this environment will change to single-column
% format so that long equations can be displayed. Use
%\begin{widetext}
%\end{widetext}

% figures should be put into the text as floats.
% Use the graphics or graphicx packages (distributed with LaTeX2e)
% and the \includegraphics macro defined in those packages.
% See the LaTeX Graphics Companion by Michel Goosens, Sebastian Rahtz,
% entire page. There is no need to do explicit centering.

% \begin{figure}
% \end{figure}

% Surround figure environment with turnpage environment for landscape
% figure
% \begin{figure}
% \end{figure}

\clearpage

\clearpage

\clearpage

\clearpage

\clearpage

\clearpage

\clearpage

\clearpage

\clearpage

\clearpage

% tables should appear as floats within the text
% The ruledtabular enviroment adds doubled rules to table and sets a
% reasonable default table settings.
% Use the table* environment to get a full-width table in two-column
% Add \usepackage{longtable} and the longtable (or longtable*}
% environment for nicely formatted long tables. Or use the the [H]
% placement option to break a long table (with less control than 
% \begin{table}%[H] add [H] placement to break table across pages
% \begin{ruledtabular}
% Lines of table here ending with \\
% \end{tabular}
% \end{ruledtabular}
% \end{table}

% Surround table environment with turnpage environment for landscape
% table
% \begin{table}
% \begin{ruledtabular}
% \end{tabular}
% \end{ruledtabular}
% \end{table}

\begin{table}%
\caption{\label{tab0}
Division of the orbital set in each symmetry. 
}
\begin{ruledtabular}
\begin{tabular}{lrrrrrrrr}
Symmetry  & $A_g$ & $B_{2u}$ & $B_{3u}$ & $B_{1g}$ &  $B_{1u}$ &
   $B_{3g}$ &  $B_{2g}$ &   $A_u$    \\
\hline
Valence   & 1-3$a_g$ & 1$b_{2u}$ &  1$b_{3u}$ &  &  1-3$b_{1u}$ &
  1$b_{3g}$  &  1$b_{2g}$ &   \\ 
Extra virtual   & 4$a_g$ & 2$b_{2u}$ &  2$b_{3u}$ & 1$b_{1g}$ & 
4$b_{1u}$ &  2$b_{3g}$ & 2$b_{2g}$  & 1$a_u$  \\
Continuum & 5-39$a_g$ & 3-35$b_{2u}$ &  3-35$b_{3u}$ & 
2-17$b_{1g}$  &  5-37$b_{1u}$ &  3-18$b_{3g}$ &  
3-18$b_{2g}$ & 2-17$a_u$    \\ 
\end{tabular}
\end{ruledtabular}
\end{table}

\clearpage

\begin{table}%
\caption{\label{tab1}
 Comparison of the vertical excitation energies.  
 The present results are shown with the previous works of 
 Gillan et al.\cite{JPhysB.29.1531}, multi reference coupled-cluster (MRCC) 
 results of Ben-Shlomo and Kaldor \cite{JChemPhys.92.3680} 
 as well as experimental values quoted in 
 Ben-Shlomo and Kaldor \cite{JChemPhys.92.3680}. 
 The unit of energy is eV.
}
\begin{ruledtabular}
\begin{tabular}{lrrrr}
State & This work & Previous R-matrix & MRCC & Experimental values\\
\hline
${X}^1 \Sigma^{+}_{g}$ &  0.00   & 0.00 & 0.00 & 0.00 \\
${A}^3 \Sigma_u^+$     &  7.89   & 7.63 & 7.56 & 7.75 \\
${B}^3 \Pi_{g}$        &  8.54   & 8.54 & 8.05 & 8.04 \\
${W}^3 \Delta_{u}$     &  9.38   & 9.11 & 8.93 & 8.88 \\
${a}^1 \Pi_{g}$        &  9.85   & 9.89 & 9.27 & 9.31 \\
${B'}^3 \Sigma^{-}_{u}$ & 10.06 & 9.83 & 9.86 & 9.67 \\
${a'}^1 \Sigma^{-}_{u}$ & 10.69 & 10.41 & 10.09 & 9.92  \\
${w}^1 \Delta_{u}$      & 11.01 & 10.74 & 10.54 & 10.27  \\
${C}^3 \Pi_{u}$         & 11.64 &       & 11.19 & 11.19 \\
\end{tabular}
\end{ruledtabular}
\end{table}

\end{document}